\documentclass[letterpaper, 10 pt, conference]{./cls/ieeeconf}
\IEEEoverridecommandlockouts
\usepackage{cite}
\usepackage{amsmath,amssymb,amsfonts}
\usepackage{algorithmic}
\usepackage{graphicx}
\usepackage{textcomp}
\usepackage{xcolor}
\usepackage{cases}
\usepackage{multirow}

\usepackage{mathtools} %for mathclap

\usepackage[ruled,norelsize]{algorithm2e}

\newtheorem{definition}{Definition}

\usepackage{./cls/Thinh_package/slashbox}
\newcommand{\be}{\begin{equation}}
\newcommand{\ee}{\end{equation}}

\newcommand{\bald}{\begin{aligned}}
\newcommand{\eald}{\end{aligned}}

\newcommand{\bbm}{\begin{bmatrix}}
\newcommand{\ebm}{\end{bmatrix}}

\newcommand{\bxi}{\boldsymbol{\xi}}
\newcommand{\bu}{\boldsymbol{u}}

\newcommand{\bx}{\boldsymbol{x}}
\newcommand{\bv}{\boldsymbol{v}}
\newcommand{\bI}{\boldsymbol{I}}
\newcommand{\R}{\mathbb{R}}

\newcommand{\bsig}{\boldsymbol{\sigma}}

\newtheorem{rem}{Remark}
\newtheorem{prop}{Proposition}
\usepackage{tikz}
\usepackage{physics}
\usepackage{amsmath}
\usepackage{tikz}
\usepackage{mathdots}
\usepackage{yhmath}
\usepackage{cancel}
\usepackage{color}
\usepackage{siunitx}
\usepackage{array}
\usepackage{multirow}
\usepackage{amssymb}
\usepackage{tabularx}
\usepackage{extarrows}
\usepackage{booktabs}
\usetikzlibrary{fadings}
\usetikzlibrary{patterns}
\usetikzlibrary{shadows.blur}
\usetikzlibrary{shapes}

\usepackage{pgfplots}
\pgfplotsset{width=10cm,compat=1.9}

\def\BibTeX{{\rm B\kern-.05em{\sc i\kern-.025em b}\kern-.08em
    T\kern-.1667em\lower.7ex\hbox{E}\kern-.125emX}}

\usepackage{pgfplots}
\usepackage[hidelinks]{hyperref} 

\pgfplotsset{compat=newest}
\pgfplotsset{plot coordinates/math parser=false}
\newlength\figureheight
\newlength\figurewidth
\usepgfplotslibrary{patchplots}
\usepackage{changepage}   % for the adjustwidth environment, add indent

\begin{document}
 \setlength{\tabcolsep}{2pt}

\title{\LARGE \bf Indoor experimental validation of MPC-based trajectory tracking \\for a quadcopter via a flat mapping approach 
}

% \author{\IEEEauthorblockN{Huu-Thinh DO}
% \IEEEauthorblockA{\textit{Univ.\ Grenoble Alpes,}
% \textit{Grenoble INP}\\
% LCIS, F-26000, Valence, France \\
% huu-thinh.do@lcis.grenoble-inp.fr}
% \and
% %%%%%%%%%%%%%%%%%%%%%%%%%%%%%%%%%%%%%%
% \IEEEauthorblockN{Ionela PRODAN}
% \IEEEauthorblockA{\textit{Univ.\ Grenoble Alpes,}
% \textit{Grenoble INP}\\
% LCIS, F-26000, Valence, France \\
% ionela.prodan@lcis.grenoble-inp.fr}
% % \and
% %%%%%%%%%%%%%%%%%%%%%%%%%%%%%%%%%%%%%%%
% % \IEEEauthorblockN{Florin STOICAN}
% % \IEEEauthorblockA{
% % \textit{Politehnica University of Bucharest}\\
% % ACSE, Bucharest, Romania \\
% % florin.stoican@upb.ro}
% % \and
% % \IEEEauthorblockN{4\textsuperscript{th} Given Name Surname}
% % \IEEEauthorblockA{\textit{dept. name of organization (of Aff.)} \\
% % \textit{name of organization (of Aff.)}\\
% % City, Country \\
% % email address or ORCID}
% % \and
% % \IEEEauthorblockN{5\textsuperscript{th} Given Name Surname}
% % \IEEEauthorblockA{\textit{dept. name of organization (of Aff.)} \\
% % \textit{name of organization (of Aff.)}\\
% % City, Country \\
% % email address or ORCID}
% % \and
% % \IEEEauthorblockN{6\textsuperscript{th} Given Name Surname}
% % \IEEEauthorblockA{\textit{dept. name of organization (of Aff.)} \\
% % \textit{name of organization (of Aff.)}\\
% % City, Country \\
% % email address or ORCID}
% }
% \title{\LARGE \bf
% Preparation of Papers for IEEE Sponsored Conferences \& Symposia*
% }
\author{Huu-Thinh DO$^{1}$ and Ionela PRODAN$^{1}$
\thanks{$^{1}${Univ. Grenoble Alpes,}
{Grenoble INP$^\dagger$},
LCIS, F-26000, Valence, France (email:{\tt\small \{huu-thinh.do,ionela.prodan\}
@lcis.grenoble-inp.fr})  
$^\dagger$ Institute of Engineering and Management Univ. Grenoble Alpes. This work is partially funded by La Région, Pack Ambition Recherche 2021 - PlanMAV and by the French National Research Agency in the framework of the ``Investissements d'avenir" program ``ANR-15-IDEX-02" and the LabEx PERSYVAL ``ANR-11-LABX-0025-01".}
}

% Univ. Grenoble Alpes, Grenoble INP$^\dagger$, LCIS,  F-$26000$, Valence, France (e-mail: ionela.prodan@lcis.grenoble-inp.fr). \\ $^\dagger$ Institute of Engineering and Management Univ. Grenoble Alpes.
\maketitle

\begin{abstract}
Differential flatness has been used to provide diffeomorphic transformations for non-linear dynamics to become a linear controllable system. This greatly simplifies the control synthesis since in the flat output space, the dynamics appear in canonical form (as chains of integrators).  The caveat is that mapping from the original to the flat output space often leads to nonlinear constraints. In particular, the alteration of the feasible input set greatly hinders the subsequent calculations. 
In this paper, we particularize the problem for the case of the quadcopter dynamics and investigate the deformed input constraint set. An optimization-based procedure will achieve a non-conservative, linear, inner-approximation of the non-convex, flat-output derived, input constraints. Consequently, a receding horizon problem (linear in the flat output space) is easily solved and, via the inverse flat mapping, provides a feasible input to the original, nonlinear, dynamics. Experimental validation and comparisons confirm the benefits of the proposed approach and show promise for other class of flat systems.
% Differential flatness has always been a significant tool to construct the diffeomorphism transforming a non-linear system into a linear controllable one together with an endogenous dynamic feedback. Ideally, without any constraints, this transformation will facilitate the control design since in the new flat output coordinates, the system is described in canonical form (or chains of integrators). However, practically, the controller construction is usually hindered by the altered constraint set via such coordinate change. In this paper, we investigate the case of the quadcopter model with the deformation of the input constraint set after such flatness-based input transformation. Then, an optimization-based approximation procedure will be introduced to achieve a non-conservative linear representation of the input in the flat space and guarantee the constraint satisfaction in the original coordinates. Next, with those stepping-stones, an MPC controller is constructed with the linearized dynamics employing a quadratic cost and linear constraints. Experimental validation is also provided to confirm the applicability of the proposed contributions. 
\end{abstract}

\begin{keywords}
Differential flatness, unmanned aerial vehicle, quadcopter, feedback linearization, model predictive control.
\end{keywords}
% \begin{IEEEkeywords}
% Differential flatness, quadcopter, feedback linearization, constraints, model predictive control.
% \end{IEEEkeywords}
\vspace{-0.2cm}
\section{Introduction}
In the context of the global contagious pandemic in recent years, 
% unmanned aerial vehicles (UAV), and particularly 
multicopters, have received remarkable attention thanks to their mobility and a wide range of applications in transportation and delivery. Although these systems have been analyzed for decades \cite{formentin2011flatness,freddi2011feedback}, the question of optimally improving the tracking performance is still open due to their strong nonlinearity and the presence of physical constraints.

To tackle the issue, while several approaches have been proposed \cite{formentin2011flatness,freddi2011feedback,nguyen2020stability}, we focus our attention on Model Predictive Control (MPC) since it has the ability of computing the optimal control inputs while ensuring constraint satisfaction. 
% Namely, this strategy allows one to obtain a model-based prediction for the system's behavior in a finite horizon, hence computing the optimal solution in real-time subject to operating constraints. 
In the literature, there are various strategies for implementing MPC in real-time with multicopters. For example, one option is to directly consider the nonlinear dynamics and design a nonlinear controller \cite{nguyen2020stability}. However, in practice, solving a nonlinear optimization problem requires a high computation power. Thus, apart from applying MPC in a nonlinear setting, the dynamics is usually governed by exploiting its model inversion given by the theory of differentially flat systems \cite{fliess1995flatness}.
Indeed, the quadcopter system is 
famously known to accept a special representation through differential flatness (i.e., all its states and inputs are algebraically expressed in terms of a flat output and a finite number of its derivatives).
Based on such property, on one hand, one method is to exploit that \textit{flat representation} to construct an integral curve (a  solution of the system's differential equation) by parameterizing the flat output in time with sufficient smoothness. Then, along such curve, the dynamics are approximated by linear time varying models 
and controlled with linear MPC \cite{prodan2015predictive}. 
Expectedly, this method provides a computational advantage thanks to the approximated linear model while the convexity of the input constraint is preserved. Yet, one shortcoming is that the model relies on an approximation, leading to the sensitivity of the controller to uncertainties. On the other hand, in the context of exact linearization, the nonlinear dynamics can be linearized in closed-loop using a linearization law taken from the relation between the flat output and the inputs \cite{nguyen2020stability, greeff2018flatness}, making the stability more straightforward to analyze. However, the drawback is that, during the linear control design in the new coordinates, the constraint set will be altered, leading to a nonlinear or even non-convex set. This phenomenon is often dealt by online prediction,
% \cite{kandler2012differential}, 
conservative offline approximation \cite{nguyen2020stability,mueller2013model} or constraint satisfying feedforward reference trajectory. 

This paper addresses the challenges and benefits of synthesizing a controller within the flat output space of a multicopter.
% using its feedback linearizable properties. 
Since in the new space coordinates, the input constraints are convoluted, we formulate a zonotope-based inner approximation of the constraints. Then a classical MPC for the linearized system is employed, which shows good performances in practice. Briefly, our main contributions are:
\begin{itemize}
    \item describe and investigate the system's constraints in the new coordinates deduced from its flatness properties;
    \item propose an optimization-based procedure to
    find a maximum inscribed inner approximation of the aforementioned altered constraint set by adopting the technique of rescaling zonotopes \cite{ioan2019navigation};
    \item construct an MPC scheme for the multicopter linear dynamics and experimentally validate it with the Crazyflie 2.1 platform in comparison with the piece-wise affine approach \cite{prodan2015predictive} and other existing results in the literature.
\end{itemize}

The remainder of the paper is structured as follows.
Section \ref{sec:sys_description} presents the system's dynamics together with its input sets, both before and after the closed-loop linearization via flatness. 
% By exploiting the convexity of the new constraint set, a zonotopic-based approximation procedure is provided in Section \ref{sec:zono_app}. 
Exploiting the convexity of the new constraints, we provide a zonotopic-based approximation procedure in Section \ref{sec:zono_app}.
The effectiveness of our approach is then experimentally validated in Section \ref{sec:exp}. Finally, Section \ref{sec:conclude} draws the conclusion and discusses future directions.

\emph{Notation:} Bold capital letters refers to the matrices with appropriate dimension. $\boldsymbol I_n$ and $\boldsymbol 0_{n}$
% and $\boldsymbol 0_n$ 
denote the identity and zero matrix of dimension $(n\times n)$, respectively. Vectors are represented by bold letters. $\text{diag}(\cdot)$ denotes the diagonal matrix created by the employed components.
 $\|\bx\|_{\boldsymbol{Q}}\triangleq\sqrt{\bx^\top \boldsymbol{Q}\bx}$. For discrete system, $\bx_k$ denotes the value of $\bx$ at time step $k$. The superscript ``$\mathrm{ref}$'' represents the reference signal (e.g, $\bx^\mathrm{ref}$). Next, 
$\mathcal{N}_m\triangleq\{1,...,m\}$ denotes the set of integers $i$ such that $1\leq i \leq m$. Furthermore, $\partial\mathcal{X}$ and $\text{int}(\mathcal{X})$ respectively denote the boundary and the interior of the set $\mathcal{X}$. Finally, $\text{Conv}\{\cdot\}$ denotes the convex hull operation.
 
  \vspace{-0.075cm}
 \section{System description and input constraints formulation in the flat output space}
  \vspace{-0.075cm}
 \label{sec:sys_description}
 In this section, we briefly present the quadcopter model together with its flat characterization.
 % which later leads to the input transformation law linearizing the model in closed-loop. 
 Next, due to the variable change, the input constraint of the quadcopter is transformed into a different non-convex set which then is replaced by a convex alternative.
 \vspace{-0.2cm}
 \subsection{Flat characterization of quadcopter model}
  \vspace{-0.1cm}
Let us recall the  quadcopter translational dynamics:
\begin{equation}
\begin{aligned}
	\bbm\ddot x \\ \ddot y \\ \ddot z \ebm&\text=\bbm 
		 T(\cos \phi \sin \theta \cos \psi\text+\sin \phi \sin \psi) \\
		 T(\cos \phi \sin \theta \sin \psi{ - }\sin \phi \cos \psi) \\
		 T\cos \phi \cos \theta -g
	\ebm
% \triangleq 	
 \text{= }
 \boldsymbol h_\psi(\bu),
\end{aligned}
	\label{eq:drone_dyna}
\end{equation} 
where $x,\, y,\, z$ are the positions of the drone, $\psi$ denotes the yaw angle, $g$ is the gravity acceleration and $\bu=[ T \,\;\phi \,\;\theta]^\top\in\R^3$ collects the inputs including the normalized thrust, roll and pitch angle, respectively. Finally,  $\mathcal{U}$ denotes the input constraint set, which is described
as:
\be 
\mathcal{U}=\{\bu :0\leq  T\leq  T_{max},\; |\phi|\leq \phi_{max},\;|\theta|\leq \theta_{max}\}
\label{eq:orginal_input_constr}
\ee
% $\mathcal{U}=\{\bu:0\leq T\leq  T_{max},\; |\phi|\leq \phi_{max},\;|\theta|\leq \theta_{max}\}$ 
where $ T_{max}$ and $\theta_{max},\phi_{max}\in ( 0;{\pi}/{2})$ are, respectively, the upper bound of $ T$ and $|\phi|,|\theta|$.

To compensate the system's nonlinearity, one typical solution is to construct its \textit{flat representation} \cite{fliess1995flatness}, i.e, parameterizing all the system's variables with a special output, called the \textit{flat output}, and its derivatives. Then, based on such model inversion, one can define a coordinate change associated with a dynamic feedback linearizing the system in closed-loop. Indeed, this quadcopter model is known to be differentially flat, and its flat representation can be expressed as\cite{nguyen2018effective}:
\begin{subequations}
	\begin{align}
% x&=\sigma_1, y=\sigma_2, z=\sigma_3,\label{eq:flat_repa}  \\
 T&=\sqrt{\ddot\sigma_1^2+\ddot\sigma_2^2+(\ddot\sigma_3+g)^2},\label{eq:flat_repb} \\
\phi&=\arcsin{\left({(\ddot\sigma_1\sin{\psi}-\ddot\sigma_2\cos{\psi})}/{ T}\right)},\label{eq:flat_repc} \\
\theta&=\arctan{\left({(\ddot\sigma_1\cos{\psi}+\ddot\sigma_2\sin{\psi})}/{(\ddot\sigma_3+g)}\right)},\label{eq:flat_repd}
	\end{align}
	\label{eq:flat_rep}
\end{subequations}
% \be
% \begin{aligned}
% x&=\sigma_1, y=\sigma_2, z=\sigma_3  \\
% T&=\sqrt{\ddot\sigma_1^2+\ddot\sigma_2^2+(\ddot\sigma_3+g)^2} \\
% \phi&=\arcsin{\left({(\ddot\sigma_1\sin{\psi}-\ddot\sigma_2\cos{\psi})}/{T}\right)} \\
% \theta&=\arctan{\left({(\ddot\sigma_1\cos{\psi}+\ddot\sigma_2\sin{\psi})}/{(\ddot\sigma_3+g)}\right)}
% \end{aligned}
% \ee 
with the flat output $\bsig=[\sigma_1,\sigma_2,\sigma_3]^\top\triangleq[x\,\,y\,\,z]^\top$. 

Next, by exploiting \eqref{eq:flat_repb}-\eqref{eq:flat_repd}, we employ an input transformation which is compactly written as $\bu=\boldsymbol \varphi_\psi(\bv)$ and detailed in \eqref{eq:linearization}. In the mapping, $\bv=[v_1,v_2,v_3]^\top$ collects the input the new coordinates called the \textit{flat output space}.
\begin{subequations}
	\begin{align}
 T&=\sqrt{v_1^2+v_2^2+(v_3+g)^2},\label{eq:linearization_a} \\
\phi&=\arcsin{\left({(v_1\sin{\psi}-v_2\cos{\psi})}/{ T}\right)},\label{eq:linearization_b} \\
\theta&=\arctan{\left({(v_1\cos{\psi}+v_2\sin{\psi})}/{(v_3+g)}\right)}.\label{eq:linearization_c}
	\end{align}
\label{eq:linearization}
\end{subequations}
Then, under the condition of $v_3\geq-g$ and the mapping \eqref{eq:linearization}, the system \eqref{eq:drone_dyna} is transformed into:
% \vspace{-0.2cm}
\be 
\ddot \bsig =\bv, \text{ with } \bsig,\bv \in\R^3.
\label{eq:fbL}
% \vspace{-0.12cm}
\ee 
Ideally, without constraints, the system can be controlled by closing the loop for the trivial system \eqref{eq:fbL}.
However, as a consequence of the input mapping \eqref{eq:linearization}, the constraint $\mathcal{U}$ in \eqref{eq:orginal_input_constr} becomes geometrically altered. Hence, it is of importance to analyze the alternation to construct a suitable controller. Indeed, hereinafter, we pave the way towards the flatness-based MPC (FB-MPC) design for the linear system \eqref{eq:fbL} by constructing the constraint set for $\bv$ in the new space. 

\begin{figure}[htbp]
    \centering
    \resizebox{0.46\textwidth}{!}{\input{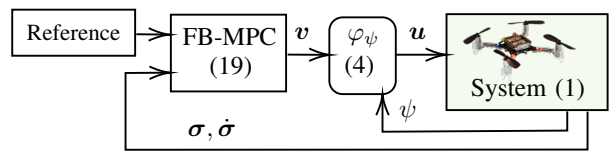}}
    \caption{Flatness-based MPC control scheme.}
    \label{fig:scheme}
\end{figure}

A general overview of the proposed control scheme for quadcopter control is in Fig. \ref{fig:scheme}. With the reference deduced from the parameterization of the flat output and the feedback signal, the FB-MPC controller computes the necessary input $\bv$ to compensate the error based on the linear model \eqref{eq:fbL} and the input constraint set in the flat output space. Then the new input $\bv$ will be mapped back to the original coordinates as $\bu$ using the transformation \eqref{eq:linearization}. Finally, the control $\bu$ is sent to the drone, ensuring the tracking performance.
\vspace{-0.3cm}
\vspace{-0.5cm}
\subsection{Input constraint characterization in the flat output space}
As aforesaid, the constraint set $\mathcal{U}$ is complicated by means of \eqref{eq:linearization}. 
Let us denote the new constraint set for $\bv$ as:
\be 
\mathcal{V}=\left\{
\bv\in\R^3\;|\;\boldsymbol \varphi_\psi(\bv)\in\mathcal{U} \text{ as in \eqref{eq:orginal_input_constr}}
\right\}.
\label{eq:constraint_in_v}
\ee 
\vspace{-0.35cm}
\begin{rem}
It is essential to point out that our motivation to construct such a set lies on the structural property of the mapping $\boldsymbol \varphi^{-1}_\psi(\bu)$. Indeed, since the function $\boldsymbol \varphi_\psi(\bv)$ is continuous and continuously invertible ($\boldsymbol \varphi^{-1}_\psi(\bu)=\boldsymbol h_\psi(\bu)$), it describes a \textit{homeomorphism} which maps the interior and boundary of a set, respectively, to those of its image.
% , i.e., $\boldsymbol \varphi^{-1}_\psi(\mathrm{int}(\mathcal{U}))=\mathrm{int}(\boldsymbol \varphi^{-1}_\psi(\mathcal{U}))$; $\boldsymbol \varphi^{-1}_\psi(\partial\mathcal{U})=\partial\boldsymbol \varphi^{-1}_\psi(\mathcal{U})$. 
Hence, under this mapping,  some geometrical properties of $\mathcal{U}$ (e.g, compactness and connectedness) are preserved in $\mathcal{V}$, encouraging us for a later-mentioned approximation.
\end{rem}
% \vspace{-0.1cm}

Regardless, it can be shown that $\mathcal{V}$ in \eqref{eq:constraint_in_v} is non-convex: the two vectors $\bv_\pm=\boldsymbol h_\psi([ T_{max},\pm\phi_{max}, \pm\theta_{max}]^\top) \in \mathcal{V}$ but $(\bv_+ + \bv_-)/2\notin \mathcal{V}$. Moreover, $\mathcal{V}$ appears to be impractical owing to its dependence on the yaw angle $\psi$, which in real applications, is certainly time-variant. For these reasons, let us consider the following subset of $\mathcal{V}$, denoted as $\Tilde{\mathcal{V}}$:
\vspace{-0.05cm}
\be 
\begin{aligned}
    \Tilde{\mathcal{V}}\triangleq\Big\{\bv\in \R^3:  \bbm {v_1^2+v_2^2+(v_3+g)^2} - T_{max}^2\\
    % \sqrt{\dfrac{v_1^{2}+v_2^{2}}{v_1^{2}+v_2^{2}+(v_3+g)^{2}}}-  \sin\epsilon_{max} 
    v_1^2+v_2^2- (v_3^2+g)^2 \tan^2\epsilon_{max} 
    \ebm    \leq 0, &\\ 
    \epsilon_{max}\triangleq\min(\theta_{max},\phi_{max}) \text{ and } v_3\geq-g\Big\}.&
    \end{aligned}
    \label{eq:constraint_convex}
\ee 
\begin{prop}
$\Tilde{\mathcal{V}}$ in \eqref{eq:constraint_convex} is convex and $\Tilde{\mathcal{V}}\subset {\mathcal{V}}$ as in \eqref{eq:constraint_in_v}. 
\label{prop:tVc}
\end{prop}

\begin{figure}[hbp]
%.\MATLAB\Quadrotor\Input_constraints_relax
    \centering
    \includegraphics[scale = 0.325]{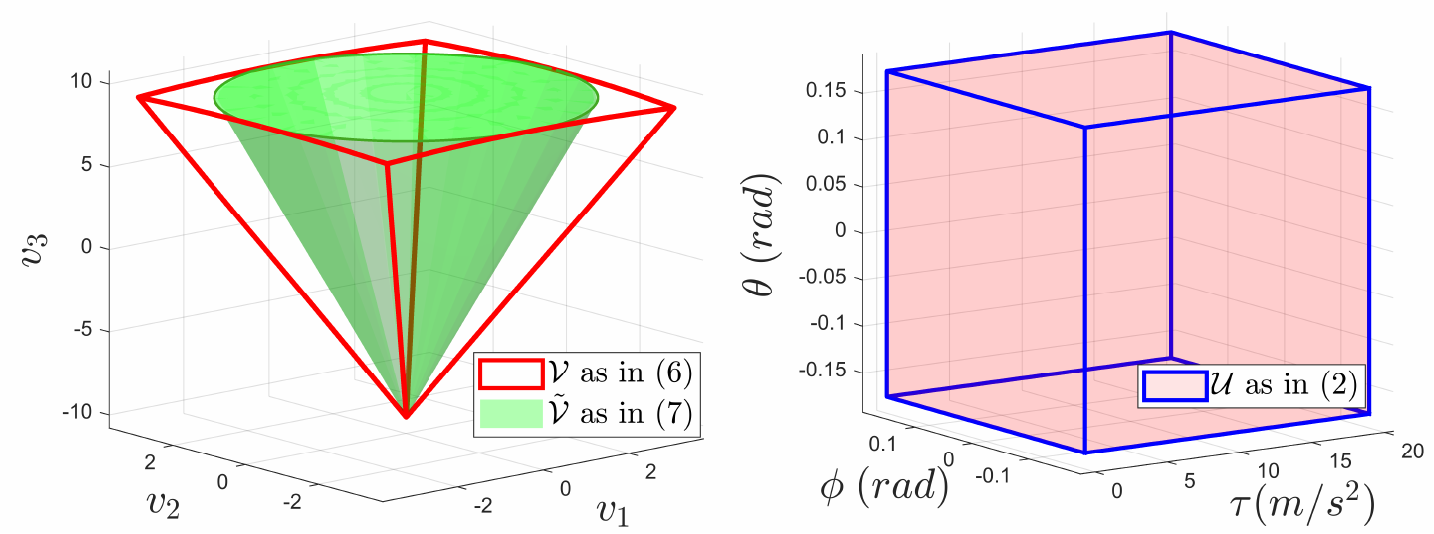}
    \caption{Constraint sets for the input $\bv$ in the flat output space (left) and the constraint set $\mathcal{U}$ in the original space (right).}
    \label{fig:Draft_V_Vtilde}
    \vspace{-0.35cm}
\end{figure}

\begin{proof}
First, to show that $\Tilde{\mathcal{V}}\subset {\mathcal{V}}$, from \eqref{eq:linearization}, by using Cauchy-Schwarz inequality, we can construct the upper bounds for the roll and pitch angles $\phi,\theta$ as follows \cite{nguyen2018effective}:
\begin{subnumcases}
{
 \label{eq:thinhset}
}
 |\sin\phi|\leq \sqrt{{(v_1^{2}+v_2^{2})}/{\left(v_1^{2}+v_2^{2}+(v_3+g)^{2}\right)}},
 & \label{eq:thinhseta} 
 \\
|\tan\theta| \leq \sqrt{{(v_1^2+v_2^2)}/{(v_3+g)^2}}.
& \label{eq:thinhsetb}
\end{subnumcases}

After some trigonometric transformations, we can see that the right-hand side of both \eqref{eq:thinhseta} and \eqref{eq:thinhsetb} represent the same angle: $\epsilon_0(\bv) \triangleq \arctan \sqrt{{(v_1^2+v_2^2)}/(v_3+g)^2}$. Hence, \eqref{eq:thinhset} yields:
% \be 
$
|\sin\phi|\leq \sin{\epsilon_0(\bv)}\text{ and } 
|\tan\theta| \leq \tan{\epsilon_0(\bv)}.$
% \label{eq:thinhset_short}
% \ee 
Then, with $|\phi|,|\theta|<\pi/2$ as in \eqref{eq:orginal_input_constr} and by imposing $\epsilon_0(\bv)\leq \epsilon_{max}$, we have $(|\phi|,|\theta|)\leq (\theta_{max},\phi_{max})$. Thus, if $\bv\in\Tilde{\mathcal{V}}$, $\bv\in {\mathcal{V}}$. 
% More details on the trigonometric transformation can be found in .
Moreover, the 
convexity of  $\Tilde{\mathcal{V}}$ can be shown by analyzing the intersection of the two convex sets: a ball of radius $ T_{max}$ and a convex cone defined by the two inequalities $v_1^2+v_2^2- (v_3^2+g)^2 \tan^2\epsilon_{max}\leq 0$, $v_3\geq -g$ (see Fig. \ref{fig:Draft_V_Vtilde}). 
\end{proof}

Up to this point, the constrained control problem is reduced to governing the linear system \eqref{eq:fbL}, under the convex constraints $\bv\in \Tilde{\mathcal{V}}$ as in \eqref{eq:constraint_convex}, depicted in Fig. \ref{fig:Draft_V_Vtilde}. However, to exploit more the advantage of this linear dynamics, it would be computationally beneficial if $\Tilde{\mathcal{V}}$ can be approximated by linear constraints, hence, reducing the complexity of the control problem. 
% In the literature, there exist several efforts to conservatively approximate the quadcopter constraints by a box-type set \cite{mueller2013model,greeff2018flatness,nguyen2020stability}. 
Thus, in the next section, by parameterizing a family of zonotopes, an optimization problem will be introduced to achieve a tractable representation for $\tilde{\mathcal{V}}$.
\vspace{-0.25cm}
 \section{Input constraints approximation \texorpdfstring{\\ in the flat output space}{}}
 \label{sec:zono_app}
With the idea of approximating the set by inflating a geometric object and achieve the largest volume inscribed, ellipsoids are typically employed owing to their volume's explicit expression\cite{boyd2004convex}. However, with ellipsoids, few advantages can be of use both geometrically and computationally if employed with MPC. Hence, let us exploit the benefits of zonotopes for which we not only have the volume's explicit formula but also obtain linear constraints. 
% alternatively, let us introduce in this section the employment of zonotopes from which we can also have an explicit formula for the volume, then by imposing such sets inside $\Tilde{\mathcal{V}}$, an optimization problem can be formulated to achieve agreeable results.
\vspace{-0.15cm}
\subsection{Zonotope parameterization}
Let us first recall the definition of a zonotope.
\begin{definition} 
In $\mathbb R^d$, given a \textit{center} point $\boldsymbol c$ and a set of $n_g$ vectors $\{\boldsymbol g_1,...,\boldsymbol g_{n_g}\}$, then $\mathcal{Z}(\boldsymbol G,\boldsymbol c)$ is called a \textit{zonotope} and can be described as \cite{mcmullen1971zonotopes}:
\be 
\mathcal{Z}(\boldsymbol G,\boldsymbol c)=\{\boldsymbol c+\textstyle\sum_{i=1}^{n_g}\beta_i\boldsymbol g_i:|\beta_i|\leq 1\},
\label{eq:zono_def}
\ee 
with $\boldsymbol G=[\boldsymbol g_1, \boldsymbol g_2,...,\boldsymbol g_{n_g}]$ gathering all the \textit{generators} $\boldsymbol g_i$. A zonotope is, indeed, a centrally symmetric polytope.
\end{definition}
Additionally, the following properties can be established.
Consider the following set $\mathbf E(\mathcal{Z}(\boldsymbol G,\boldsymbol c)) $ defined as:
\begin{align}
\mathbf E (\mathcal{Z}(\boldsymbol G,\boldsymbol c))\triangleq\big\{\boldsymbol c+\textstyle\sum_{i=1}^{n_g}\alpha_i\boldsymbol g_i,
% \nonumber \\
% &\;\;\;\;
|\alpha_i|=1
% \subset\R^{n_g}
\big\}. \label{eq:list_ver}
\end{align}
Then, the set $\mathbf E(\mathcal{Z}(\boldsymbol G,\boldsymbol c)) $ in \eqref{eq:list_ver} is a \textit{finite}
subset of $\mathcal{Z}(\boldsymbol G,\boldsymbol c)$ and contains all of its vertices \cite{mcmullen1971zonotopes}. Consequently, for a convex set $\mathcal{X}$, the inclusion $\mathcal{Z}(\boldsymbol G,\boldsymbol c)\subseteq \mathcal{X}$ holds if and only if $\mathbf E(\mathcal{Z}(\boldsymbol G,\boldsymbol c)) \subset\mathcal{X}$. 
 Furthermore, as  proposed in \cite{ioan2019navigation}, consider a family of parameterized zonotopes: 
 \be 
 \mathcal{Z}(\boldsymbol G\Delta,\boldsymbol c)=\{\boldsymbol c+\textstyle\sum_{i=1}^{n_g}\beta_i\delta_i\boldsymbol g_i:|\beta_i|\leq 1\},
 \label{eq:zono_parameterized}
 \ee 
where $\boldsymbol G$ is a given generator matrix and $\Delta=\text{diag}(\boldsymbol \delta)$ denotes a diagonal matrix whose diagonal is collected in $\boldsymbol \delta=[\delta_1,...,\delta_{n_g}]$. 
 Then, its volume is explicitly written as:
%  \footnote{I have no idea how to drag $\delta_k$ closer to $\prod$ in \eqref{eq:vol}}:
\be 
\mathcal{C }(\boldsymbol\delta)=\sum_{1\leq k_1<...<k_d\leq n_g} \left| \det(\boldsymbol G^{k_1...k_d})\right|\;\;\;\, \prod_{\mathclap{{k\in\{k_1,...,k_d\}}}}\delta_k
\label{eq:vol}
\ee 
where $\boldsymbol G^{k_1...k_d}\in \mathbb R^{d\times d}$ denotes the matrix formed by stacking the $k_l$-th column, $l\in\mathcal{N}_d$, of $\boldsymbol G$ together. 

Using the above tools, we propose in the next subsection a zonotopic inner-approximation for the set $\tilde{\mathcal{V}}$ in \eqref{eq:constraint_convex}.
% With these apparatus, in the next subsection, let us define a maximization problem for a zonotope inscribed in a convex set, particularly in this case as $\tilde{\mathcal{V}}$.
% \vspace{-0.1cm}
% \begin{figure}[htbp]
%     \centering
%     \includegraphics[scale=0.4]{pics/ECC23/zono_ex.pdf}
%     \caption{Approximation of the set $\mathcal{X}=\{(x_1,x_2)\in\R^2:x_2\geq 2x_1-2,x_1^2+x_2^2\leq 25\}$ using \eqref{eq:opt_new_drone} with different choices of $\bx_0$ and $G=\bI_2$. The green area denotes the convex hull of all resulted sets.}
%     \label{fig:zono_ex}
% \end{figure}
\vspace{-0.05cm}
\subsection{Constraints approximation in the flat output space}
\vspace{-0.025cm}
In here, we construct an optimization problem to find the largest zonotope from the family of $\mathcal{Z}(\boldsymbol G\Delta,\boldsymbol c)$ as in \eqref{eq:zono_parameterized} constrained in the convex set $\tilde{\mathcal{V}}\subset\R^3$:
 \begin{subequations}
	\begin{align}
		&(\boldsymbol \delta,\boldsymbol c)^*     =\underset{                     
		\boldsymbol \delta,\boldsymbol c 
		}{\arg\operatorname{max}}\;
		\mathcal{C }(\boldsymbol \delta)
		,\label{eq:opt_new_drone_a}\\
	\text{s.t } &  \bv\in \tilde{\mathcal{V}},\forall \bv \in \mathbf E(\mathcal{Z}(\boldsymbol G\Delta,\boldsymbol c)),
\label{eq:opt_new_drone_b}\\
&\bv^\mathrm{int} \in \mathcal{Z}(\boldsymbol G\Delta,\boldsymbol c) ,\label{eq:opt_new_drone_c}
	\end{align}
	\label{eq:opt_new_drone}
	\vspace{-0.03cm}
\end{subequations}
where $\mathcal{Z}(\boldsymbol G,\boldsymbol c)\subset\R^3$  denotes the zonotope formed by the given $n_g$ generators in $\boldsymbol G\in\R^{3\times n_g}$ centered at $\boldsymbol c$; $\boldsymbol \delta=[\delta_1,...,\delta_{n_g}]\in\R^{n_g}$ is a scaling factor magnifying the original zonotope $\mathcal{Z}(\boldsymbol G,\boldsymbol c)$. $\mathcal{C }(\boldsymbol \delta)$ denotes the volume of $\mathcal{Z}(\boldsymbol G\Delta,\boldsymbol c)$ computed as in \eqref{eq:vol} with $\Delta\triangleq\text{diag}(\boldsymbol \delta)$. Next, $\mathbf E (\mathcal{Z}(\boldsymbol G\Delta,\boldsymbol c))\subset \mathcal{Z}(\boldsymbol G\Delta,\boldsymbol c)$ is \textit{finite} and
% is the \textit{finite} subset of $\mathcal{Z}(\boldsymbol G\Delta,\boldsymbol c)$ 
contains all vertices of $\mathcal{Z}(\boldsymbol G\Delta,\boldsymbol c)$, and can be enumerated as in \eqref{eq:list_ver}. This 
condition \eqref{eq:opt_new_drone_b} ensures 
% that the resulted zonotope will be contained in $\tilde{\mathcal{V}}$.  
the inclusion $\mathcal{Z}(\boldsymbol G\Delta,\boldsymbol c)\subset\tilde{\mathcal{V}}$ as previously discussed.
Finally, $\bv^\mathrm{int}\in\mathrm{int}(\tilde{\mathcal{V}})$ is a user-defined point towards which the resulting set is allowed to expand. Specifically, since zonotopes are symmetric, it is impossible for them to expand freely inside $\tilde{\mathcal{V}}$. Therefore, progressively imposing the condition \eqref{eq:opt_new_drone_c} with different $\bv^\mathrm{int}$ helps us obtain several zonotopes reaching to specific ``corners" of $\tilde{\mathcal{V}}$. Then the final approximated set is achieved as the convex hull of all the zonotopes deduced from those choices of  $\bv^\mathrm{int}$.

For instance, let us denote $\mathcal{I}$ the sets of choice containing some $\bv^\mathrm{int}\in \text{int}(\tilde{\mathcal{V}})$. One candidate can be enumerated as in \eqref{eq:N0} which contains $N_0+1$ points taken between two extreme ones of $\tilde{\mathcal{V}}$: $[0,0,-g]^\top$ and $[0,0, T_{max}-g]^\top$:
 \be
\begin{aligned}
\mathcal{I}=\{\bv^\mathrm{int}_0,..., \bv^\mathrm{int}_{N_0}\}& 
% \\
% \mathcal{I}=\big\{ \bv_0\in\R^3: \bv_{0k}=[0,0,1]^\top((1-{k}/{N_0}) T_{max}-g), &\\
% k\in\{0,...,N_0\}\big\}.&
\end{aligned}
\label{eq:N0}
\ee
with $\bv^\mathrm{int}_k=[0,0,1]^\top((1-{k}/{N_0}) T_{max}-g)$.
Then, for each $0\leq k \leq N_0$ as in \eqref{eq:N0}, we obtain from \eqref{eq:opt_new_drone} a parameterized zonotope $\mathcal{Z}(\boldsymbol G\Delta,\boldsymbol c)$, assigned as $\mathcal{S}_v^k$. Finally, 
the resulting approximation set is computed as:
\be 
\mathcal{S}_v=\text{Conv}\left\{
\mathcal{S}_v^0,...,\mathcal{S}_v^{N_0}\right \}.
\label{eq:convH}
\ee 
The procedure is summarized in Algorithm \ref{alg:zono_approx}.
\vspace{-0.05cm}
\begin{algorithm}[hbtp!]
\caption{$\tilde{\mathcal{V}} $ approximation procedure.}
\label{alg:zono_approx}
\KwIn{Generator $\boldsymbol G$ as in \eqref{eq:opt_new_drone} and $\mathcal{I}$ as in \eqref{eq:N0}.}
\KwOut{The approximation set $\mathcal{S}_v$ as in \eqref{eq:convH} of  $\tilde{\mathcal{V}}$.}
\For{$k=0$  \KwTo $N_0$ }
{
$\bv^\mathrm{int}_k\gets[0,0,1]^\top((1-{k}/{N_0}) T_{max}-g)$\;
Solve the optimization problem \eqref{eq:opt_new_drone} for $(\boldsymbol \delta,\boldsymbol c)$\;
$\mathcal{S}_v^k \gets \mathcal{Z}(\boldsymbol G\Delta,\boldsymbol c)$ with $\Delta=\text{diag}(\boldsymbol \delta)$\;
}
$\mathcal{S}_v\gets \text{Conv}\left\{
\mathcal{S}_v^0,...,\mathcal{S}_v^{N_0} 
\right\} $  as in \eqref{eq:convH}.
\end{algorithm}
% \vspace{-0.05cm}
\vspace{-0.05cm}
\begin{rem}
Although the semi-definite condition \eqref{eq:opt_new_drone_b} is straightforward to construct,
% , 
% compared to the semi-infinite programming formula proposed in \cite{faiz2001trajectory} with the same context with flatness.
% of exploiting flatness properties for enlarging polytopes ($ \bv\in\tilde{\mathcal{V}}\,\forall \bv\in \mathcal{Z}(\boldsymbol G\Delta,\boldsymbol c)$). 
% However, 
it comes with a shortcoming: the computational cost 
% (the number of constraints in \eqref{eq:opt_new_drone_b}) 
rises exponentially with the number of generators, $n_g$, since the number of elements in $\mathbf E(\mathcal{Z}(\boldsymbol G\Delta,\boldsymbol c))$  is $2^{n_g}$. This drawback is indeed burdensome, especially for the approximation of high dimensional sets, because one needs a sufficiently large number of generators to have a good basis zonotope $\mathcal{Z}(\boldsymbol G,\boldsymbol c)$  to be scaled with $\boldsymbol \delta$.
\end{rem}
 \subsection{Simulation result}
\begin{figure}[htbp]
 \vspace{-0.15cm}
    \centering
    \includegraphics[scale = 0.42]{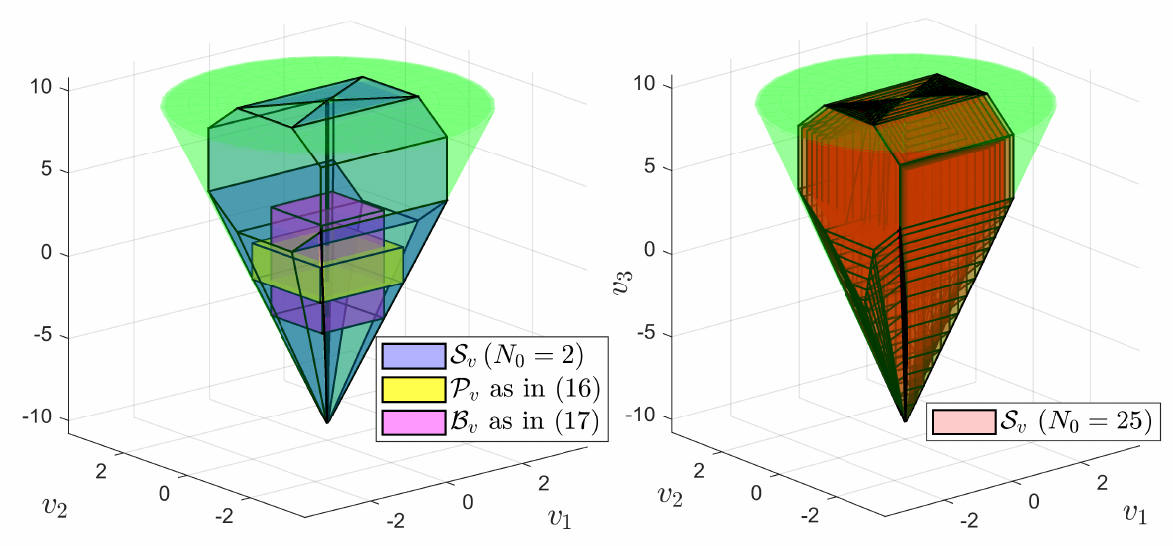}
    \caption{Approximated constraint sets for the new input $\bv$.}
    \label{fig:First_and_last_Zono_pp_NT_eq_box}
\vspace{-0.35cm}
\end{figure}

 In this part, we discuss the simulation result of applying Algorithm \ref{alg:zono_approx} for $\tilde{\mathcal{V}}$ as in \eqref{eq:constraint_convex} with $\mathcal{I}$ enumerated as in \eqref{eq:N0}.
 For illustration, we examine two scenarios with $N_0=2$ and $N_0=25$ for \eqref{eq:N0}. The result, its specification and parameters setup are provided in Fig. \ref{fig:First_and_last_Zono_pp_NT_eq_box}
 % Fig. \ref{fig:Vol_ineq} 
 and Table \ref{tab:First_and_last_Zono_pp}, respectively.
%  \vspace{-0.4cm}
% \vspace{-0.25cm}
\begin{table}[htbp]
     \centering
     \caption{Parameters setup for $\Tilde{\mathcal{V}}$ approximation}
     \begin{tabular}{|c|c|}
     \hline
         Symbols & Values   \\ \hline \hline
         $g$, $ T_{max}$  &  9.81; 19.62    $m/s^2$\\ \hline
        % $ T_{max}$  &  2g  $m/s^2$   \\ \hline
                % $ T_{max}$  &   $2  g\;m/s^2$   \\ \hline

        $\theta_{max}, \phi_{max}$  &  0.1745  $rad$ $(10^o)$   \\ \hline
        $\boldsymbol G$  &  $\bbm 1 & 0 & 0 & 0 & 0
        \\ 0&1&0&1&-1\\
        0&0&1&2&2\ebm$   \\ \hline
        $\bar v_{1p}=\bar v_{2p}=\bar v_{3p}$  &  1.0875\\ \hline
        $[\bar v_{1b},\bar v_{2b},\bar v_{3b}]$  &  $[0.815,0.815,3.270]$\\ \hline
     \end{tabular}
     \label{tab:First_and_last_Zono_pp}
      \vspace{-0.172cm}
 \end{table}

For comparison, we examine other constraint sets for quadcopter in the literature, which are constructed as follows.
\begin{itemize}
    \item In \cite{nguyen2020stability}, a box-type 
    subset of $\Tilde{\mathcal{V}}$ was introduced as:
    \be
    \mathcal P_v=\{\bv\in \R^3: |v_i|\leq \bar v_{ip}, i\in\{1,2,3\}\}
    \ee 
    with the constant $\bar v_{ip}$ satisfying the conditions:
$$
\begin{cases}
\bar v_{3p}<g; \,\bar v_{1p}^2+\bar v_{2p}^2\leq(-\bar v_{3p}+g)^2\tan^2\epsilon_{max}&\\
\sqrt{\bar v_{1p}^2+\bar v_{2p}^2+(\bar v_{3p}+g)^2}\leq  T_{max}.&
\end{cases}
$$
\item The approximated origin-centered constraint set, in \cite{mueller2013model,greeff2018flatness}, described as:
\be \mathcal{B}_v=\{\bv\in \R^3: |v_i|\leq \bar v_{ib}, i\in\{1,2,3\}\} 
\ee
can be similarly found by employing \eqref{eq:opt_new_drone} with $\boldsymbol G=\bI_3$ and $\bv^\mathrm{int}=[0,0,0]^\top$.
\end{itemize}
 The illustration of the aforementioned sets are also shown in Fig. \ref{fig:First_and_last_Zono_pp_NT_eq_box} with their numerical values given in Table \ref{tab:First_and_last_Zono_pp} and \ref{tab:result_spec}.
  \begin{table}[htp]
  \vspace{-0.1cm}
     \centering
     \caption{Result specifications for $\Tilde{\mathcal{V}}$ approximation}
     \begin{tabular}{|l|c|c|c|c|}
     \hline
          & $\mathcal{S}_v,N_0=2$  &$\mathcal{S}_v,N_0=25$  & $\mathcal{P}_v$& $\mathcal{B}_v$\\ \hline \hline
        Volume& 122.57  &  125.27   &10.29&17.39\\ \hline
      No. of vertices&  28    &226&8&8\\ \hline
        No. of inequalities&  20  &216& 6&6\\ \hline
     \end{tabular}
     \label{tab:result_spec}
     \vspace{-0.05cm}
 \end{table}
% \vspace{-0.12cm}
% Vol_ineq.pdf
%  \begin{figure}[htbp]
%     \centering
%     \includegraphics[scale = 0.5]{pics/ECC23/Vol_ineq.pdf}
%     \caption{Number of inequalities (blue) and volume of $\mathcal{S}_v$ (red) with respect to $N_0$ as in \eqref{eq:N0}.}
%     \label{fig:Vol_ineq}
% \end{figure}
 
As depicted in Fig. \ref{fig:First_and_last_Zono_pp_NT_eq_box}, our approach provides an improved approximation for the input $\bv$, with respect to the literature. Prior to this point, the ingredients for an MPC design in the new coordinates are ready. More precisely, the system now can be governed by controlling its image in the flat output space as in \eqref{eq:fbL} with the corresponding input constraint $\mathcal{S}_v$ resulted from Algorithm \ref{alg:zono_approx}. 
Therefore, next, we validate our results via different tests within the MPC settings.
%  \begin{figure*}[ht]
%      \centering
%      \includegraphics[scale=0.5]{pics/ECC23/track_error4ref_ver2.pdf}
%      \caption{Crazyflie tracking error with flatness-based MPC and PWA-MPC (4 references) with $e_q=q_{\mathrm{ref}}-q,\;q\in\{x,y,z\}$}
%      \label{fig:trackerror_v2}
%  \end{figure*}
\vspace{-0.2cm}
 \section{MPC design with experimental validation}
 \label{sec:exp}
To demonstrate the practical viability of our approach,
% as in the schema delineated in Fig. \ref{fig:scheme},
 we first present the MPC synthesis utilizing the linear model \eqref{eq:fbL} in the new coordinates, with the corresponding linear constraint set $\mathcal{S}_v$ as in \eqref{eq:convH}.
 % With this FB-MPC setup, a control $\bv$ will be generated in the flat output space to compensate the tracking error. 
 % This input then will be transformed back to the original space as $\bu$ and applied to the system. 
 The application is conducted using the Crazyflie 2.1 nano-drone with various flight scenarios.
 \vspace{-0.2cm}
 \subsection{MPC setup}
 To embed the model \eqref{eq:fbL} and the constraints $\mathcal{S}_v$ in the MPC framework, we first proceed with the discretization as follows. Let $\bxi=[\sigma_1,\sigma_2,\sigma_3,\dot\sigma_1,\dot\sigma_2,\dot\sigma_3]^\top$ denotes the state vector of the system \eqref{eq:fbL}. Then applying Runge-Kutta (4th order) discretization method 
 to \eqref{eq:drone_dyna} yields:
 \be 
 \bxi_{k+1}=\boldsymbol A\bxi_{k} + \boldsymbol B\boldsymbol h_\psi(\bu_k)\triangleq \boldsymbol f_d(\bxi_k,\bu_k),
 \label{eq:discrete}
 \ee 
 with 
 $\boldsymbol A=\begin{bsmallmatrix}\bI_3 &t_s \bI_3\\ \boldsymbol 0_3 & \bI_3 \end{bsmallmatrix}$,
 % $A=\begin{bmatrix}\bI_3 &t_s \bI_3\\ \boldsymbol 0_3 & \bI_3 \end{bmatrix}$,
 $\boldsymbol B=[t_s^2\bI_3/2,\;t_s\bI_3]^\top$ and the sampling time $t_s$. Next, let us introduce the following controllers employed for the validation and comparison.
 
     \emph{Flatness-based MPC (FB-MPC):} with $\bu_k=\boldsymbol \varphi_\psi(\bv_k)$ as in \eqref{eq:linearization}, the system now can be controlled with the linear dynamics: 
     $  \bxi_{k+1}=\boldsymbol A\bxi_k+\boldsymbol B\bv_k
     $, subject to the new input constraint $\bv_k\in\mathcal{S}_v$ (with $N_0=2$). We solve the following online problem over the prediction horizon $N_p$ steps:
%      \be
% \begin{aligned}
% &\underset{\bv_{k},...,\bv_{k+N_p-1}}{\arg \min } \textstyle\sum_{i=0}^{N_p-1} \Big(\left\|\bxi_{i+k}-\bxi_{i+k}^\mathrm{ref}\right\|_{\boldsymbol Q}^{2}   \\[-1.2ex]
% &\;\;\;\;\;\;\;\;\;\;\;\;\;\;\;\;\;\;\;\;\;\;\;\;\;\;\;\;\;\;\;\;\;\;\;+\left\|\bv_{i+k}-\bv_{i+k}^\mathrm{ref}\right\|_{\boldsymbol R}^{2}\Big)
% \end{aligned}
% \label{eq:opti_prob_FB}
% \ee 
     \be
\underset{\mathclap{\bv_{k},{...},\bv_{k\text+N_p-1}}}{\arg \min \;\;\;\;\;} \sum_{i=0}^{N_p-1} \|\bxi_{i+k}-\bxi_{i+k}^\mathrm{ref}\|_{\boldsymbol Q}^{2}+\|\bv_{i+k}-\bv_{i+k}^\mathrm{ref}\|_{\boldsymbol R}^{2}
\label{eq:opti_prob_FB}
\ee 
\vspace{-0.1cm}
$$\text{s.t: }
\begin{cases}
\bxi_{i+k+1}=\boldsymbol A\bxi_{i+k}+\boldsymbol B\bv_{i+k},
&\\
\bv_{i+k} \in {\mathcal{S}_v},i\in\{0,...,N_p-1\}
&
\end{cases} 
\vspace{-0.3cm}$$ 
with $\bxi_{k}^\mathrm{ref},\bv_{k}^\mathrm{ref}$ gathering the reference signal of the state $\bxi$ and input $\bv$ at time step $k$. Then the first value of the solution sequence (i.e, $\bv_k$) will be used to compute the real control $\bu_k=\boldsymbol \varphi_\psi(\bv_k)$, which then is applied to the quadcopter.

\emph{Piece-Wise Affine MPC (PWA-MPC)}:  For comparison, this method is adopted from \cite{prodan2015predictive} with the linearization of the model via Taylor series and the linear MPC setup. 
% Concisely, the system dynamics will be approximated via Taylor series, which then is integrated into the MPC setup with the linear constraint $\mathcal{U}$ as in \eqref{eq:orginal_input_constr}. 
Details on the implementation are given in the Appendix.
\begin{rem}
In the flat output space, by using our convex approximation of the feasible domain, the MPC design employs both a linear model and linear constraints. Hence, within this framework, other properties (e.g, stability, robustness), either in discrete or continuous time, become more accessible for investigation \cite{mayne2006robust,blanchini2003suboptimal}. Note that these advantages do not exist for all flat systems, especially for those with the flat output different from the states (or output) of interest \cite{zafeiratou2020meshed, do2021analysis}, hence, complicating the theoretical guarantees when switching from one space to the other.
\end{rem}
\vspace{-0.05cm}
\subsection{Scenarios of trajectories}
\begin{figure}[htbp]
    \vspace{-0.1cm}
     \centering
     \includegraphics[scale=0.55]{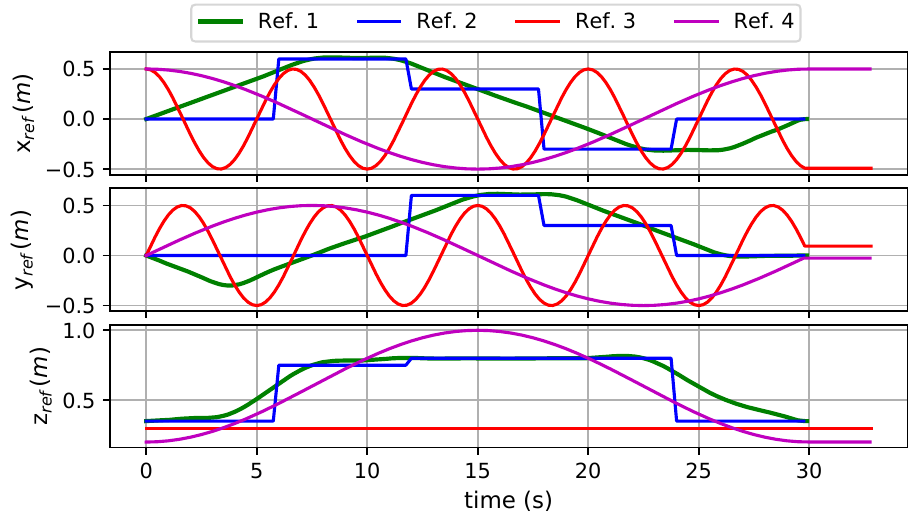}
     \caption{Four proposed time-parameterized references.}
     \label{fig:traj4}
      \vspace{-0.24cm}
 \end{figure}

To identify benefits and drawbacks of the two methods, we provide some scenarios of trajectories. In those trajectories, the flat output $\bsig$ in \eqref{eq:fbL} will be parameterized in time, which leads to the complete nominal reference of the system thanks to the representation \eqref{eq:flat_rep}. The trajectories are described as:
 \begin{itemize}
     \item Ref. 1: This reference is adopted from \cite{prodan2019necessary,do2021analysis}, with the energetically optimal B-spline parameterized curve (order 8) passing the way-points $w_k$ at $t_k\,(s)$:
\end{itemize}
     $$
     \begin{aligned}
      w_k\in\left\{
      \begin{aligned}
          (0;0; 3.5); (3;-3 ;4); (6;0;7.5);&\\
         (6;3 ;8) ;(3;6 ;8);( 0;6 ;8);&\\
          (-3;3 ;8);(-3;0 ;5);(0;0 ;3.5)&
      \end{aligned}
      \right\}\times 10 \;(cm)&\\
      t_k=(k-1)\times 30/8 \,(s), k\in\{1,2,...,9\}.&
          \end{aligned}
     $$
     % \begin{align}
%  & \begin{aligned}
%      w_k\in
%                 \Big\{\begin{bsmallmatrix}
%                 0\\0 \\3.5
%                 \end{bsmallmatrix},
%                 \begin{bsmallmatrix}
%                 3\\-3 \\4
%                 \end{bsmallmatrix},
%                 \begin{bsmallmatrix}
%                 6\\0\\7.5
%                 \end{bsmallmatrix},
%     %   \qquad \qquad 
%        \begin{bsmallmatrix}
%                 6\\3 \\8
%                 \end{bsmallmatrix},
%                 \begin{bsmallmatrix}
%                 3\\6 \\8
%                 \end{bsmallmatrix},
%                 \begin{bsmallmatrix}
%                 0\\6 \\8
%                 \end{bsmallmatrix}
%                 &
%                 \\
%                 \begin{bsmallmatrix}
%                 -3\\3 \\8
%                 \end{bsmallmatrix},
%                 \begin{bsmallmatrix}
%                 -3\\0 \\5
%                 \end{bsmallmatrix},
%                 \begin{bsmallmatrix}
%                 0\\0 \\3.5
%                 \end{bsmallmatrix}\Big\}\times 10^{-1}
%                 (m)
%   \end{aligned}& \nonumber \\
% &t_k=(k-1)\times 30/8 \,(s), k\in\{1,2,...,9\}.\nonumber
% \end{align}
\vspace{-0.25cm}
\begin{adjustwidth}{0.7cm}{}
With this method, the curve's parameters are chosen so that all the states and inputs respect their constraints, giving a favorable reference to them to follow.
\end{adjustwidth} 
 \begin{itemize}
\item Ref. 2: To make the reference more aggressive, set points are given under sequences of step functions.
% , hence reducing the amount of information given to the controllers. 

\item Ref. 3: Next, we adopt the circular trajectory in \cite{greeff2018flatness} as:
% \be 
$$
\begin{aligned}
&\sigma_{1\,\mathrm{ref}}(t)=0.5\cos\omega t, \sigma_{2\,\mathrm{ref}}(t)=0.5\sin\omega t\\
&\sigma_{3\,\mathrm{ref}}(t)=0.3(m), \omega=0.3\pi.
\end{aligned}
$$

\item Ref. 4: Finally, we adapt the arbitrary sinusoidal trajectory given in \cite{garcia2017modeling}, describing as:
% \be
$$
\begin{aligned}
&\sigma_{1\,\mathrm{ref}}(t)=0.5\cos\omega t, \sigma_{2\,\mathrm{ref}}(t)=0.5\sin\omega t\\
&\sigma_{3\,\mathrm{ref}}(t)=0.5\sin 0.5 \omega t + 0.5(m), \omega=\pi/15.
\end{aligned}
$$
% \ee 
 \end{itemize}
 Illustration of the reference trajectories are given in Fig. \ref{fig:traj4}. 
%  \vspace{-0.2cm}
  \subsection{Experimental results and discussions}
 The experiments are conducted with 8 \textit{Qualisys} motion capture cameras to have an accurate estimation of drone's position. The control signal $\bu$ is computed in a station computer, then applied to the drone by sending the desired control $ T,\phi,\theta,\psi$ via the Crazyflie PA radio USB dongle. During the experiment, the desired $\psi$ angle was set as $0$. The experiments' parameters are listed in Table \ref{tab:paras} while the \href{https://youtu.be/1a1K6R6__3s}{video} is available at: \text{\url{https://youtu.be/1a1K6R6__3s}}.
 
\begin{figure}[ht]
    \vspace{-0.1cm}
     \centering
     \includegraphics[scale=0.52]{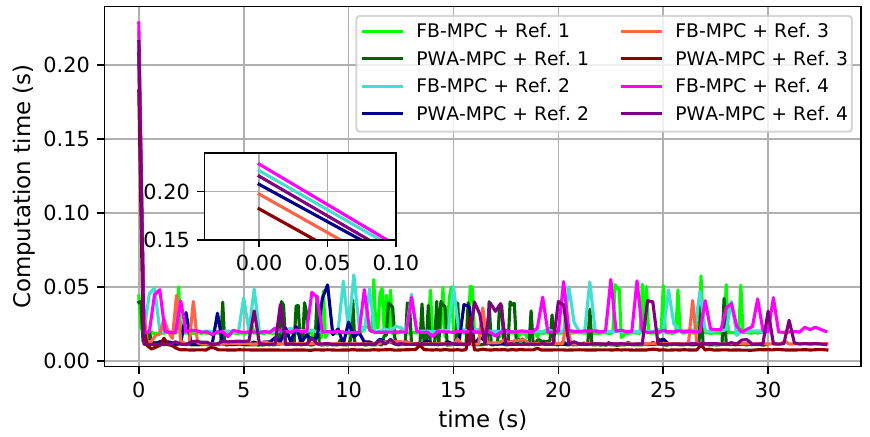}
     \caption{Computation time used for different references.}
     \label{fig:Comptime_4ref}
     \vspace{-0.25cm}
 \end{figure}

 Computationally, the sampling times were chosen according to the execution time with different trajectories. It is noticeable that with the well constructed Ref. 1, the input references $\bv_{k}^\mathrm{ref}, \bu_{k}^\mathrm{ref}$ can be nominally defined
 for the system's dynamics, hence speeding up the search for the optimal solutions in both methods. Consequently, the sampling time can be chosen only as $t_s=0.1s$, while, the remaining three trajectories demand much higher time for the initial search, resulting in larger sampling time ($\geq 0.2s$, see Fig. \ref{fig:Comptime_4ref}).
\begin{table}[hbtp]
  \centering
  \caption{Control parameters for the proposed scenarios}
    \begin{tabular}{|p{0.6cm}|l|p{2.352cm}|l|c|c|}
    \hline
         &      & \multicolumn{1}{l|}{$\boldsymbol Q$} & \multicolumn{1}{l|}{$\boldsymbol R$}  & $t_s (s)$ &$N_p$\\
    \hline
    \multirow{2}[1]{*}{Ref.1} & FB-MPC   & $\text{diag}(35\bI_2,50,5\bI_3)$     & $\bI_3$ & \multirow{2}[1]{*}{0.1} & \multirow{2}[1]{*}{20}\\
\cline{2-4}         & PWA-MPC  & $\text{diag}(35\bI_2,50,5\bI_3)$     &  $\text{diag}(5,75\bI_2)$ &&\\
    \hline
    \multirow{2}[1]{*}{Ref.2} & FB-MPC   & $\text{diag}(50\bI_3,5\bI_3)$      &  $5\bI_3$ &\multirow{2}[1]{*}{0.25}&\multirow{2}[1]{*}{20}\\
\cline{2-4}         & PWA-MPC  &    $\text{diag}(50\bI_3,5\bI_3)$   & $\text{diag}(5,75\bI_2)$ & &\\
    \hline
    \multirow{2}[1]{*}{Ref.3} & FB-MPC   &   $\text{diag}(180\bI_3,10\bI_3)$    &   $5\bI_3$ &\multirow{2}[1]{*}{0.2}&\multirow{2}[1]{*}{10}\\
\cline{2-4}         & PWA-MPC  &  $\text{diag}(50\bI_3,5\bI_3)$     & $\text{diag}(5,80\bI_2)$ & &\\
    \hline
    \multirow{2}[1]{*}{Ref.4} & FB-MPC   &  $\text{diag}(90\bI_3,5\bI_3)$     &  $5\bI_3$  &\multirow{2}[1]{*}{0.25}&\multirow{2}[1]{*}{20} \\
\cline{2-4}         & PWA-MPC  &    $\text{diag}(35\bI_2,50,5\bI_3)$   & $\text{diag}(5,75\bI_2)$ & &\\
    \hline
    \end{tabular}%
  \label{tab:paras}%
  \vspace{-0.15cm}
\end{table}%
 
\begin{figure}[htbp]
  \vspace{-0.05cm}
     \centering
     \includegraphics[scale=0.5]{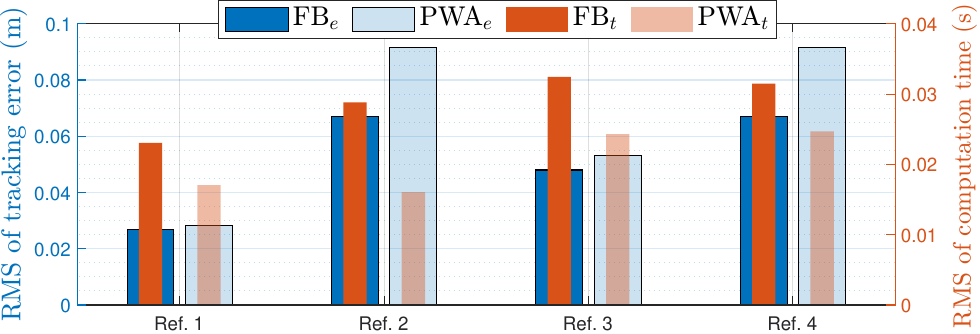}
   \caption{RMS of tracking errors and computation time of the two controllers with different types of references (distinguished, respectively, by the subscript $e$ and $t$ under the name of the corresponding controller).
   % (the values of tracking errors and computation times are distinguished, respectively, by the subscript $e$ and $t$ under the name of the corresponding controller).
   }
     \label{fig:BarRMS}
     \vspace{-0.2cm}
 \end{figure}

\begin{figure}[ht]
     \centering
     \includegraphics[scale=0.5]{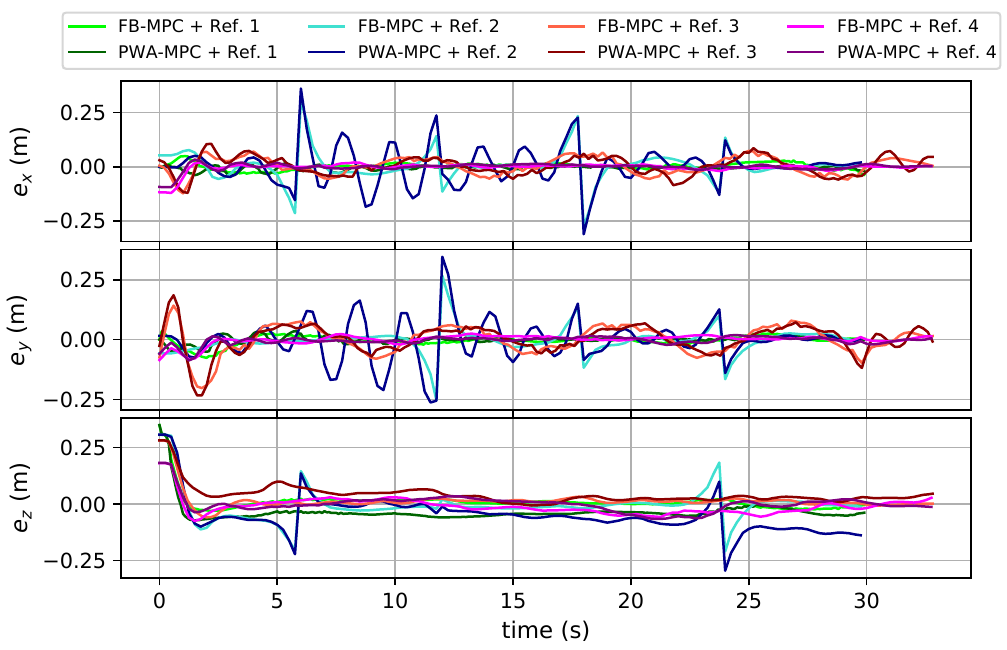}
     \caption{Crazyflie tracking error with flatness-based MPC and PWA-MPC (4 references) with $e_q\triangleq q^{\mathrm{ref}}-q,\;q\in\{x,y,z\}$.}
     \label{fig:trackerror_v2}
     % \vspace{-0.025cm}
     
 \end{figure}

In terms of performance, Fig. \ref{fig:BarRMS} and \ref{fig:trackerror_v2} show the root-mean-square (RMS) and the tracking errors of the two controllers, respectively, in the four references. Expectedly, although requiring more computation time, the FB-MPC can be considered better while being put next to the well-known PWA-MPC with centimeters of tracking error. 
 
In details, one shortcoming of FB-MPC is that it demands slightly more execution time than the PWA-MPC in practice. This can be explained by showing the complexity of the optimization problems. Particularly, both the PWA-MPC and FB-MPC as in \eqref{eq:opti_prob_FB} are quadratic programming problems. However, the constraint set $\mathcal{S}_v$ is computationally complex, compared to $\mathcal{U}$ with more vertices or inequalities. The effect can also be seen in Fig. \ref{fig:BarRMS} with a roughly constant gap in computation time necessary for the two methods. 

\begin{figure}[ht]
   \vspace{-0.05cm}
     \centering
     \includegraphics[scale=0.42]{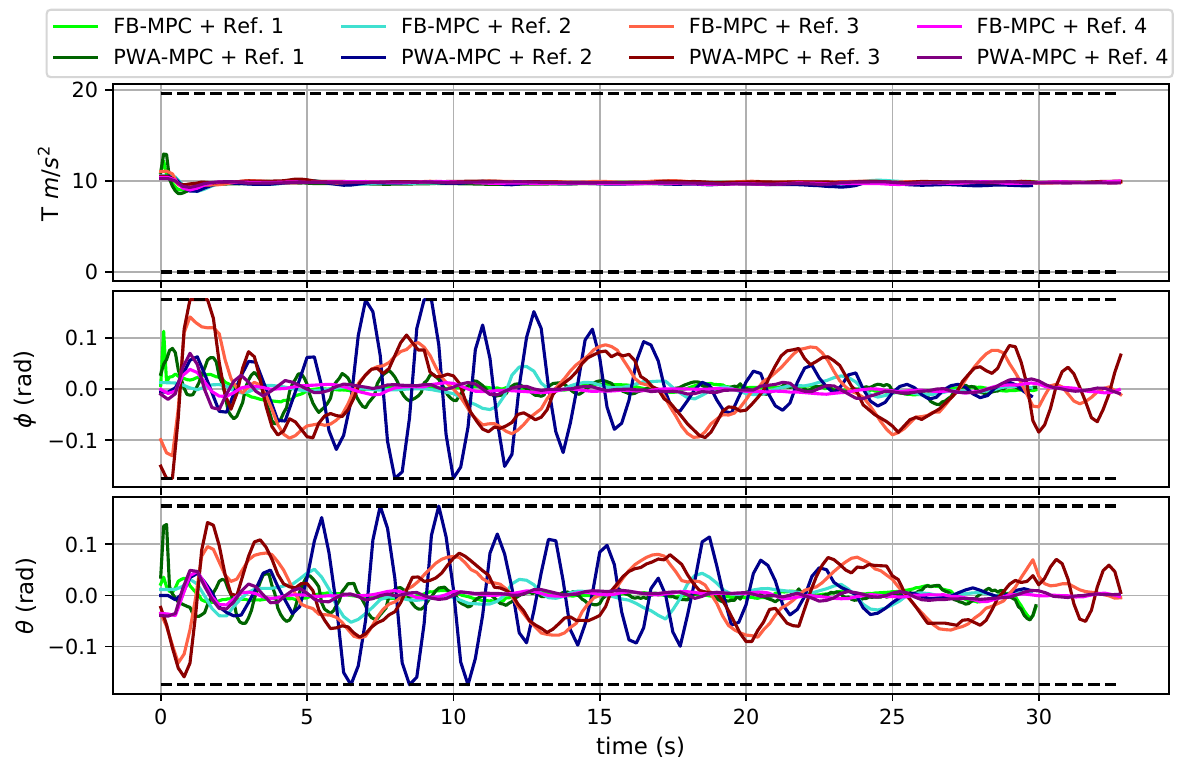}
     \caption{Input signals with their constraints (black dashed line).}
     \label{fig:Imput_sig_v2}
     \vspace{-0.3cm}
 \end{figure}

Yet, since PWA-MPC depends on the approximation of the model along the trajectory, its efficiency is reliant on the reference's quality, hence making the method more vulnerable to uncertainty than our proposed FB-MPC. Indeed, while with Ref. 1, both controllers achieve fairly good tracking (See Fig. \ref{fig:BarRMS}),
with Ref. 2, large oscillations in tracking error are observed with PWA-MPC compared to that of the FB-MPC (see Fig. \ref{fig:trackerror_v2}). Moreover, despite being constructed via an approximated input constraint, the FB-MPC always shows an equivalently reliable performance without saturating the input, in comparison with the PWA-MPC, as in Fig. \ref{fig:Imput_sig_v2}.

Furthermore, due to the fact that there is no approximation in our model, the performance in the proposed FB-MPC surpasses its approximation-based contestant in \cite{garcia2017modeling} with Ref. 3. Finally, although being constructed in the similar framework of flatness-based MPC, with Ref. 4, our improved performance is apparent thanks to the less conservative constraint set $\mathcal{S}_v$ as opposed to the box-type set in \cite{greeff2018flatness}.
 % \vfill
% \clearpage

 \section{Conclusion}
 \label{sec:conclude}
 This paper presented a reliable FB-MPC design for the quadcopter system by introducing an efficient approximation for the feasible domain in the flat output space, where the system is linearized in closed-loop. The validation demonstrates the advantages of the contributions compared to related works conducted in the literature.
 % either by approximating the nonlinear dynamics or the constraints in the flat output space. 
 % In short, by replacing the original nonlinear dynamics, with a linear one in the flat output coordinates, the quadcopter control design is significantly facilitated. 
 As future work, we attempt to adapt the procedure to other classes of flat systems, where the constraints are more geometrically distorted by the flatness-based coordinate change.

% \clearpage
 \appendix
\label{appendix:PWA}
We adapt the implementation of PWA-MPC from \cite{prodan2015predictive} as follows. 
% First, it is required to construct a trajectory respecting the system's dynamics. Then, a
Along the system's refernce trajectory, we choose $N_l$ points around which the dynamics is approximated by using Taylor expansion as:
\be
\bxi_{k+1}=\boldsymbol A_j\bxi_{k}+\boldsymbol B_j\bu_k+\boldsymbol r_j
\ee 
with $\boldsymbol f_d$ in \eqref{eq:discrete}, $\boldsymbol A_j=\left.{\partial\boldsymbol  f_d}/{\partial\bxi}\right|_{\bxi_j,\bu_j}$, $\boldsymbol B_j=\left.{\partial \boldsymbol f_d}/{\partial\bxi}\right|_{\bxi_j,\bu_j}$ and $\boldsymbol r_j=\boldsymbol f_d(\bxi_j,\bu_j)-\boldsymbol A_j\bxi_j-\boldsymbol B_j\bu_j$ while $\bxi_j,\bu_j$ respectively denote the $j$-th state and input value in the collection of $N_l$ points equidistantly chronologically sampled from the nominal trajectory. During the implementation, $\boldsymbol A_j,\boldsymbol B_j$ and $\boldsymbol R_j$ are flexibly chosen according to the drone's closest point. Hence, the online optimization problem is expressed as:
%      \be
% \begin{aligned}
% &\underset{\bu_{k},...,\bu_{k+N_p-1}}{\arg \min } \textstyle\sum_{i=0}^{N_p-1} \Big(\left\|\bxi_{i+k}-\bxi_{i+k}^\mathrm{ref}\right\|_{\boldsymbol Q}^{2}   \\[-1.2ex]
% &\;\;\;\;\;\;\;\;\;\;\;\;\;\;\;\;\;\;\;\;\;\;\;\;\;\;\;\;\;\;\;\;\;\;\;+\left\|\bu_{i+k}-\bu_{i+k}^\mathrm{ref}\right\|_{\boldsymbol R}^{2}\Big)
% \end{aligned}
% \label{eq:opti_prob_PWA}
% \ee 
     \be
\underset{\mathclap{\bu_{k},...,\bu_{k\text+N_p-1}}}{\arg \min\;\;\;\;\;\,} \sum_{i=0}^{N_p-1} \|\bxi_{i+k}-\bxi_{i+k}^\mathrm{ref}\|_{\boldsymbol Q}^{2} 
+
\|\bu_{i+k}-\bu_{i+k}^\mathrm{ref}\|_{\boldsymbol R}^{2}
\label{eq:opti_prob_PWA}
\ee
$$\text{s.t :}
\begin{cases}
\bx_{i+k+1}=\boldsymbol A_j\bxi_{i+k}+\boldsymbol B_j\bu_{i+k} + \boldsymbol r_j\\
\bu_{i+k} \in {\mathcal{U}}, \, i\in\{0,1,...,N_p-1\}
\end{cases}\vspace{-0.2cm}
$$
with $\bxi_{k}^\mathrm{ref},\bv_{k}^\mathrm{ref}$ denoting the reference for $\bxi_k$ and $\bu_k$.

% https://www.dynsyslab.org/wp-content/papercite-data/pdf/greeff-iros18.pdf
% \section{Reference*}
% \clearpage
%\bibliographystyle{IEEEtran}
%\bibliography{./Bibs/ECC2023}
\end{document}